\documentclass[a4paper]{jpconf}
\usepackage{graphicx}

\begin{document}
\title{The Belle II experiment: fundamental physics at the flavor frontier
}

\author{Ivan Heredia de la Cruz\\
\normalfont{\small{\textsl{for the Belle II Collaboration}}}
}

\address{CONACYT -- Physics Department, Centro de Investigaci\'on y de Estudios Avanzados del IPN (CINVESTAV-IPN) -- Mexico City, Mexico
}

\ead{iheredia@fis.cinvestav.mx}

\begin{abstract}
After the major success of  $B$-factories to establish the CKM mechanism and its proven potential to search for new physics, the Belle~II experiment  will continue exploring the physics  at the flavor frontier over the next years.  Belle~II  will collect 50 times more data than its predecessor, Belle, and allow for various precision measurements and searches of rare decays and particles. This paper introduces the $B$-factory concept and the flavor frontier approach to search for new physics. It then describes the SuperKEKB accelerator and the Belle~II detector, as well as some of the physics that will be analyzed in Belle~II, concluding with the experiment status and schedule.  
\end{abstract}

\section{Introduction}
The first generation of $B$-factories, PEP-II at SLAC and KEKB at KEK, provided $e^+e^-$ 
collisions at the BaBar 
and Belle 
experiments during the last decade. In particular, KEKB reached the world highest instantaneous luminosity of about $2 \times 10^{34}$~cm$^{-2}$s$^{-1}$. Collisions took place at centre-of-mass energies equal to the mass of the $\Upsilon(nS)$, mainly at $m_{\Upsilon(4S)} = 10.58$~GeV, and also at ``off-resonance" energies. BaBar and Belle collected an integrated luminosity of about 0.5 and 1~ab$^{-1}$, respectively. In Belle, 1~ab$^{-1}$ corresponds to nearly 772 million $B\bar{B}$  and 900 million $\tau \bar{\tau}$ pairs. 

The discovery potential of $B$-factories is beyond all possible doubt. For instance, Belle discovered charge-parity violation (CPV) in the $B$ meson system~\cite{CPVdiscovery1,CPVdiscovery2} and found direct CPV in  $B \rightarrow K\pi$ decays~\cite{DirectCPV1,DirectCPV2}. Belle also measured mixing in charm~\cite{Dmixing} and observed several $B$ and $\tau$ rare decays (e.g., see section~\ref{LFV}), as well as many  predicted and unexpected new states, such as the $X(3872)$~\cite{X3872} and the tetraquark candidate $Z(4430)^+$~\cite{Z4430}. 

\begin{figure}[h]
\begin{center}
\includegraphics[width=20pc]{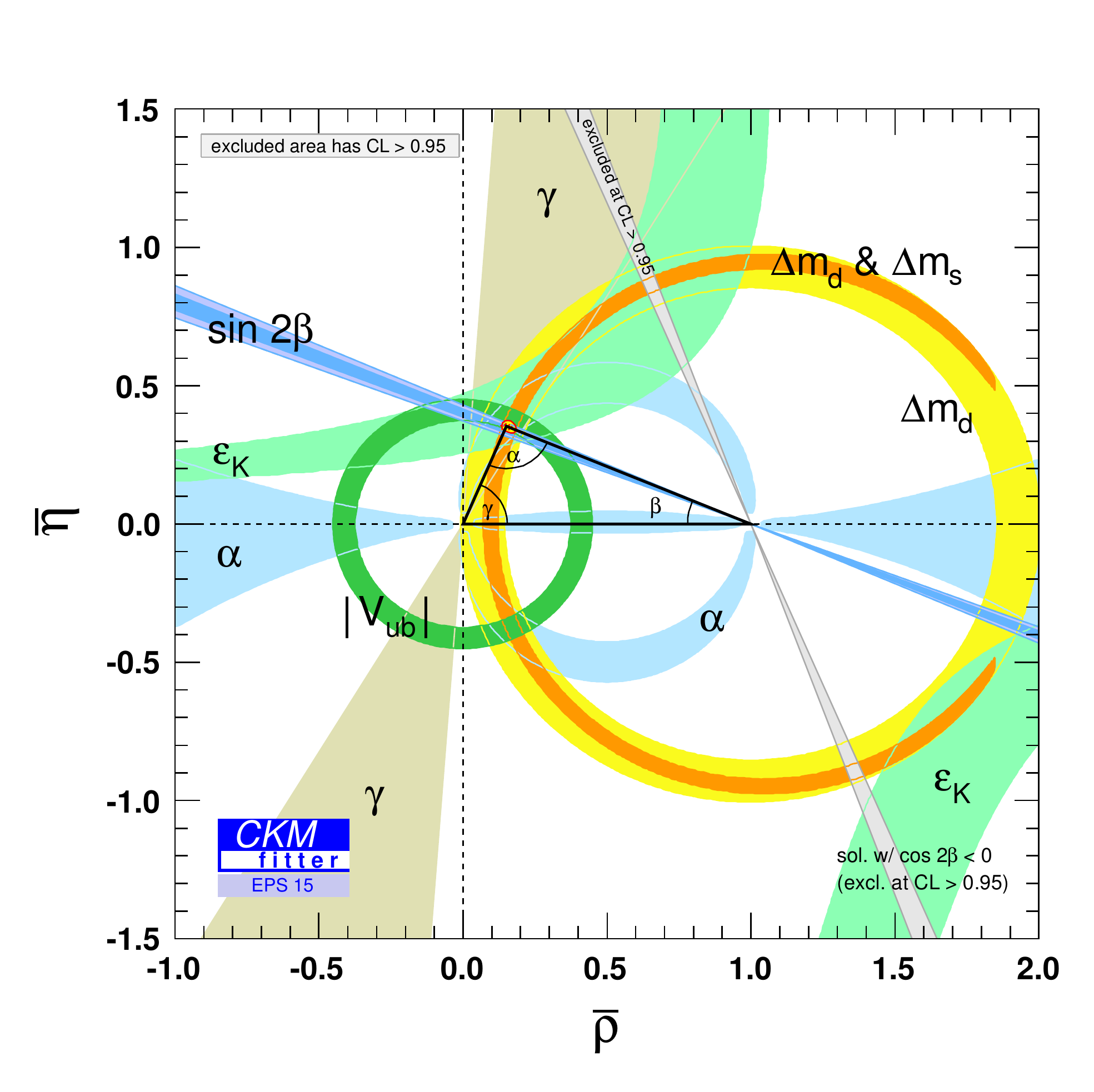}\hspace{2pc}%
\begin{minipage}[b]{14pc}\caption{\label{UT}Constrains on the CKM $(\bar{\rho}, \bar{\eta})$ plane from  the global SM CKM fit~\cite{CKMFitter}.  \vspace{0.7pc} }
\end{minipage}
\end{center}
\end{figure}

One of the greatest achievements of $B$-factories was 
the experimental confirmation of the CKM mechanism, which led to the Nobel Prize in Physics in 2008 awarded to M. Kobayashi and T. Maskawa~\cite{nobel2008}. This result, normally summarized in the so called unitary triangles  as the one in figure~\ref{UT}, shows the impressive agreement of the Standard Model (SM) global fit. 

In the light of so many impressive achievements, a second generation of $B$-factories is eagerly awaited. 
SuperKEKB and Belle II will supersede KEKB and Belle to extend the exploration of  fundamental physics at the flavor frontier.

\section{The flavor frontier}
There exist two approaches at particle colliders to search for new physics (NP), namely, the energy and flavor (or intensity) frontiers. The first produces directly new particles and  is only limited by the beam energy. The second can reveal beyond SM virtual particles appearing in loop diagrams through precise measurements of well known and rare processes. CMS and ATLAS at the LHC, for example, explore the energy frontier using 13 TeV centre-of-mass energy $pp$ collisions, while $B$-factories (and also LHCb) can test indirectly a mass scale up to 100 TeV within the flavor frontier. If there is NP at a scale of a few TeV, effects will emerge in $B$, $D$ and $\tau$ decays and, thus, will be measured by the LHC experiments. If not, flavor measurements will provide unique ways to find NP. This complementarity is crucial to understanding the need to upgrade  Belle to Belle~II, and KEK to SuperKEKB. The well defined initial state energy in Belle~II allows also for  efficient detection of inclusive decay modes (e.g. $b \rightarrow s \ell \ell$) and final states with neutral particles and missing energy (e.g. $B \rightarrow \tau \nu$), which are  complementary features   to the LHCb experiment's specialty of detecting decays to charged final states (e.g. $B \rightarrow \mu^+ \mu^-$).

A new generation of (super) $B$-factories will  make it possible to measure with higher precision almost all the CKM matrix elements through $B$ and $D$ decays. They will also be key to testing beyond SM models
by measuring or looking for mixing, CPV, lepton flavor and number violation (LFV and LNV) and rare decays of $B$, $D$  and $\tau$ particles. 
Furthermore, it is possible to search directly for light particles such as sterile neutrinos, dark photons, light Higgs bosons and axions, and to gain a better understanding of quantum chromodynamics (QCD)  by a more complete characterization of open charm, light meson and baryon states, and quarkonium and quarkonium-like exotic states.

\begin{figure}[h]
\begin{center}
\begin{minipage}{19.5pc}\vspace{1.5pc}
\includegraphics[width=20pc]{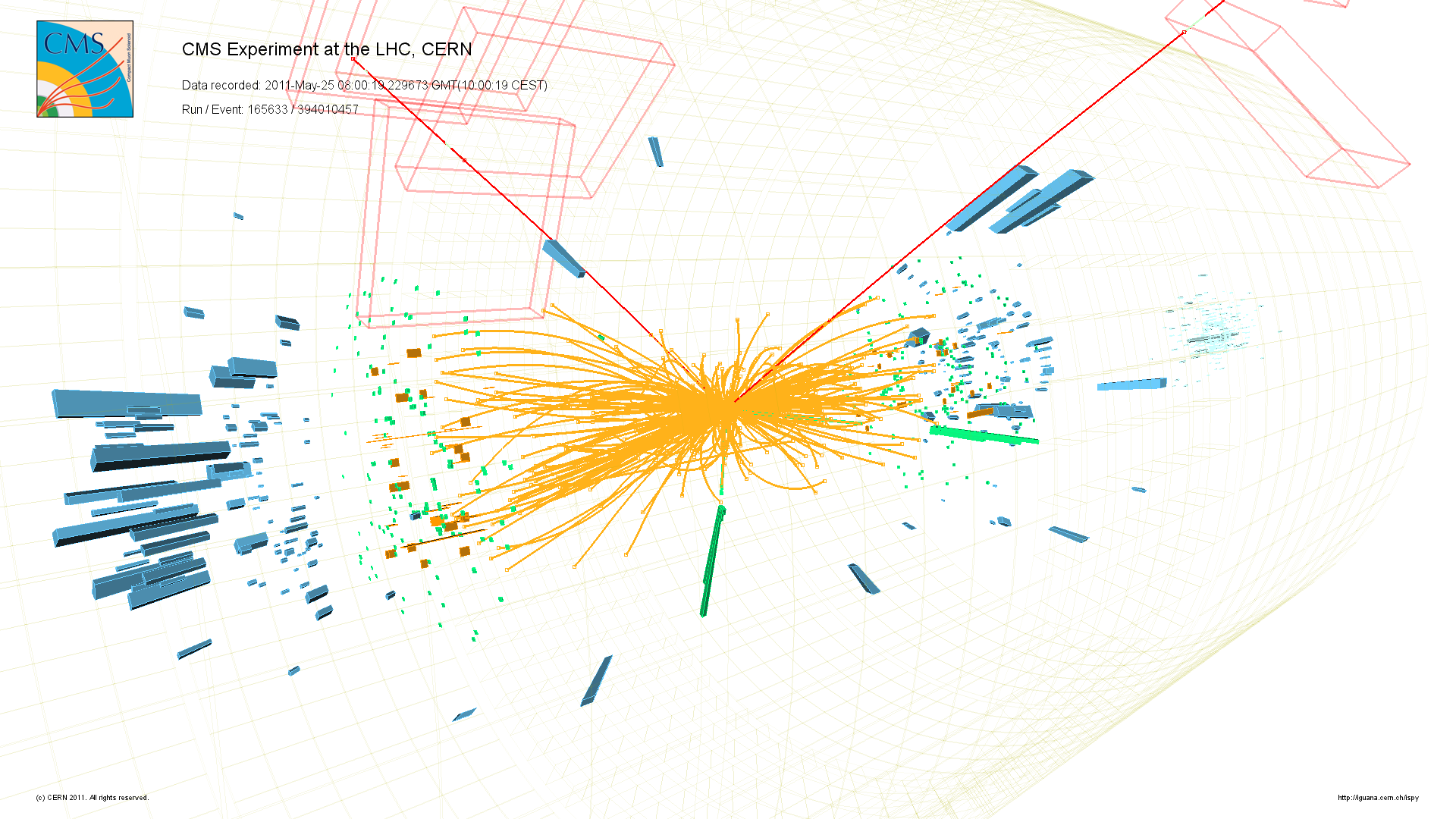}\vspace{1pc}
\caption{\label{collisionsLHC}High-multiplicity environment observed in $pp$ collisions. Event recorded by the CMS experiment~\cite{CMScollision}.}
\end{minipage}\hspace{2pc}%
\begin{minipage}{15.5pc}
\centering \includegraphics[width=14.5pc]{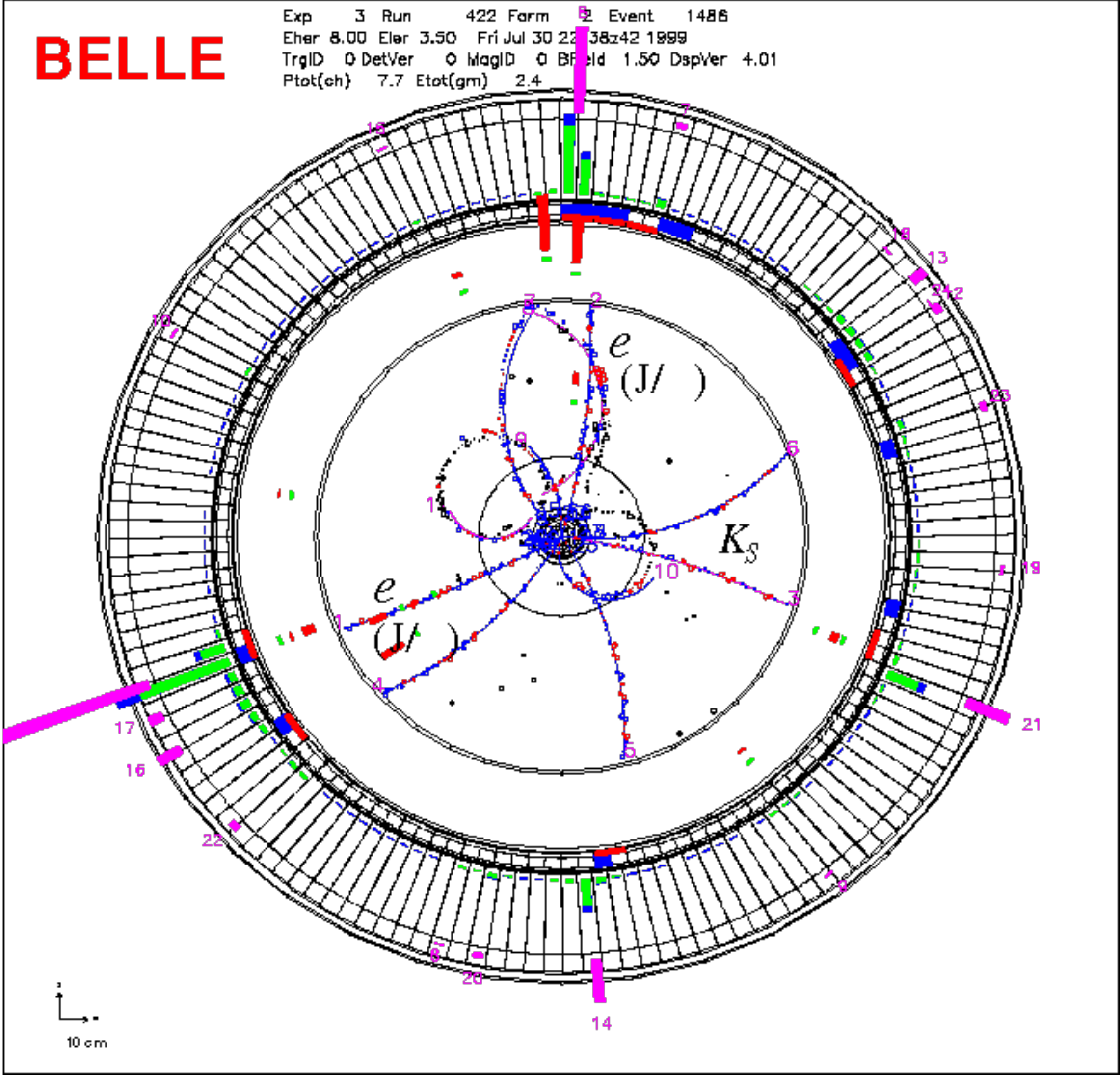}
\caption{\label{collisionsBfact}Low-multiplicity environment from $e^+e^-$ collisions. Event recorded by the Belle experiment~\cite{Bellecollision}.}
\end{minipage} 
\end{center}
\end{figure}

There are unique features in $B$-factories which make them attractive and a real alternative to the energy frontier approach. The most obvious is the much lower background  produced at lepton colliders with respect to hadron colliders, implying lower track multiplicity and detector occupancy, and in turn  resulting in higher $B$, $D$ and $\tau$ reconstruction efficiencies and easier detection of neutral particles. This point is illustrated in figures~\ref{collisionsLHC} and \ref{collisionsBfact}. Moreover, there is no need of a strict event filtering (trigger selection) that can bias a measurement and lead to intricate corrections or further systematic uncertainties. The asymmetric beam energies permit a more precise study of time dependent CPV, while  the tuning of the beams to specific $\Upsilon$ resonances results in a straightforward  bottomonium exploration. 
Another nice feature is the ability to identify a signal decay ($B_{sig}$) as the complement of the decay products of the other $B$ meson ($B_{tag}$) in the event.

\section{SuperKEKB and Belle II}
\begin{figure}[h]
\begin{center}
\begin{minipage}{17.5pc}
\includegraphics[width=17.5pc]{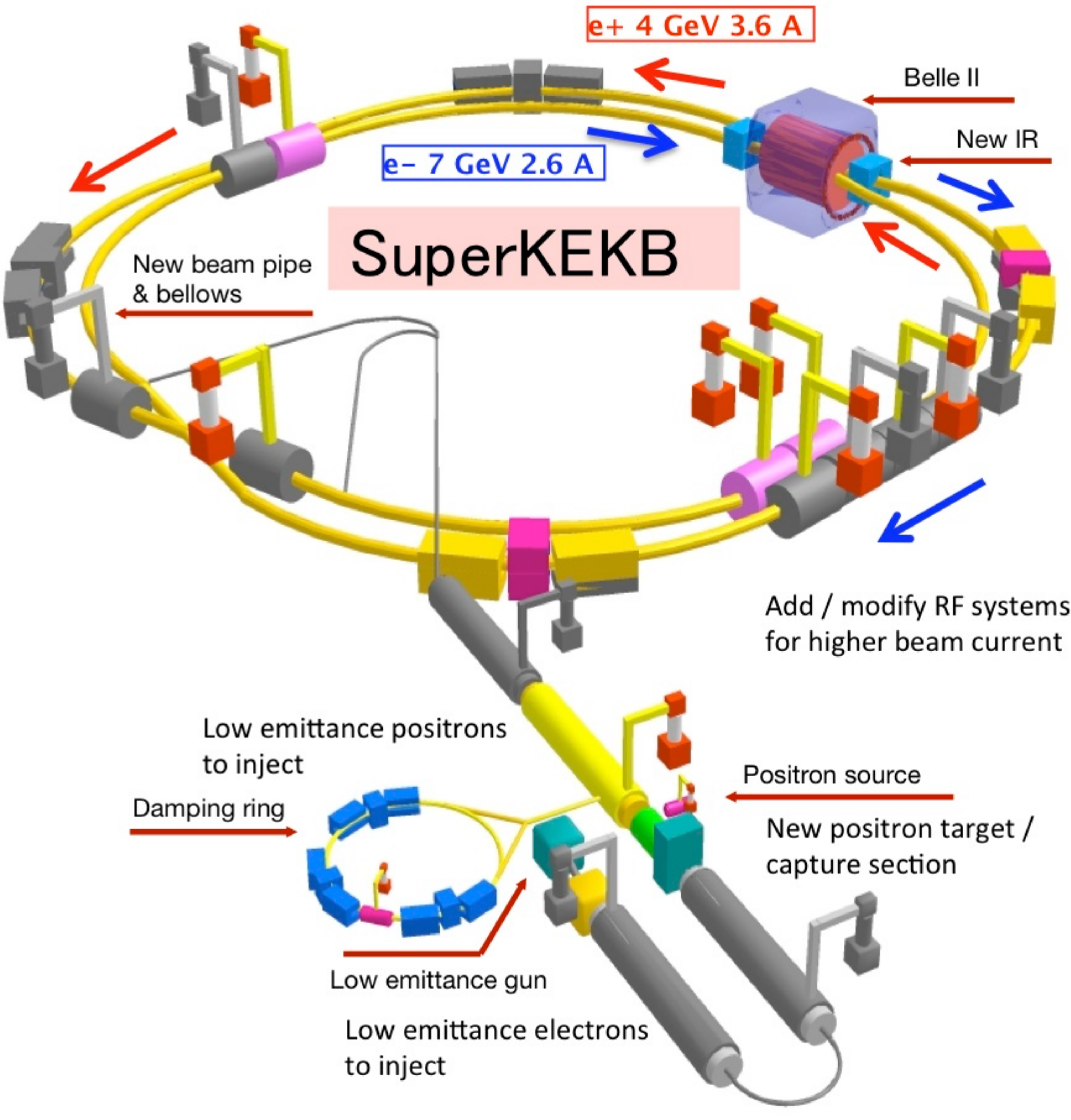}
\caption{\label{skekb}The SuperKEKB collider.}
\end{minipage}\hspace{2pc}%
\begin{minipage}{17.5pc}\vspace{1.5pc}
\includegraphics[width=17.5pc]{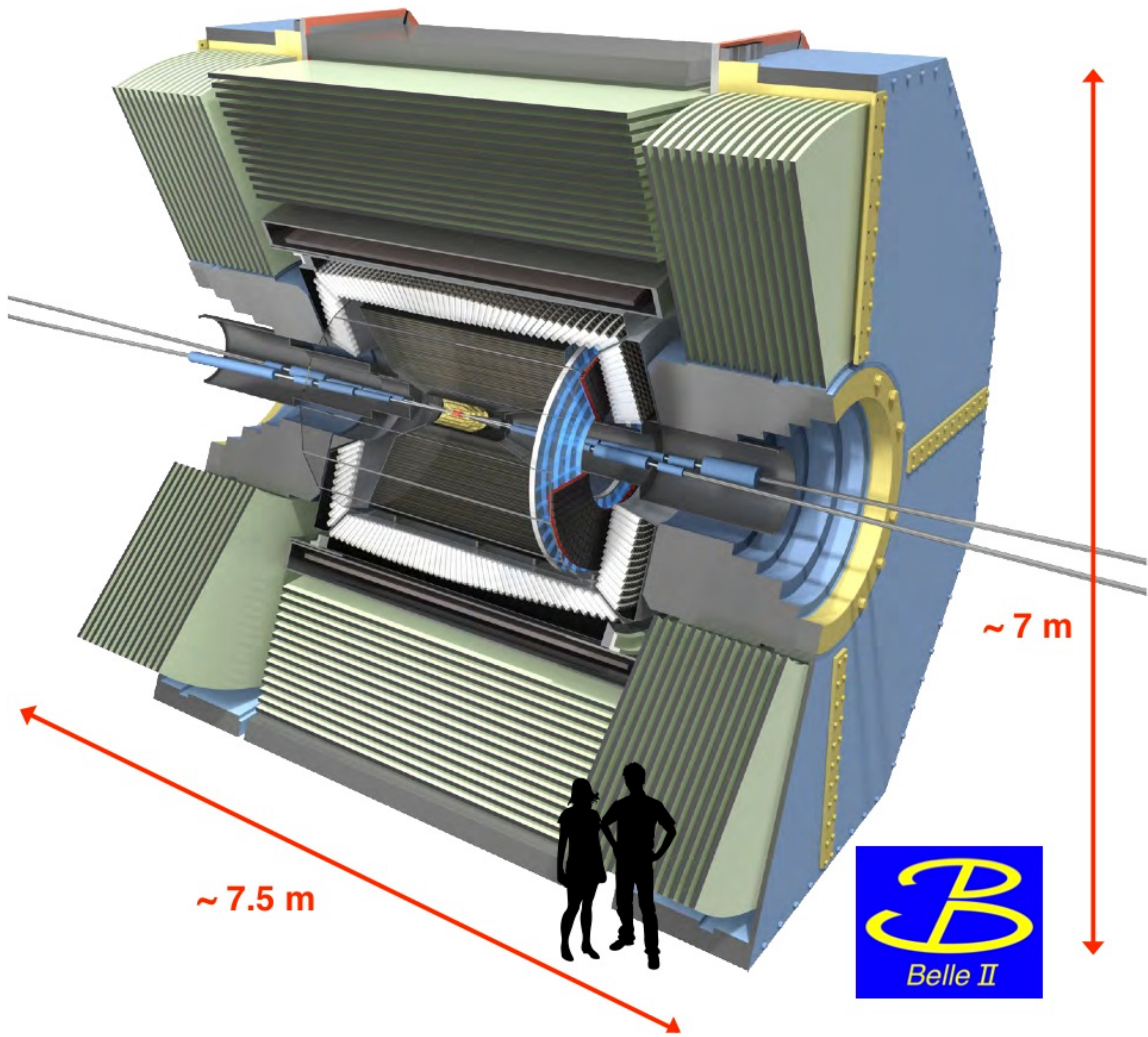}\vspace{1pc}
\caption{\label{belle2}The Belle~II detector.}
\end{minipage}
\end{center}
\end{figure}

The SuperKEKB collider and Belle II detector are shown in figures~\ref{skekb} and~\ref{belle2}, and described below.

\subsection{SuperKEKB collider}
SuperKEKB is located at KEK (The High Energy Accelerator Research Organization) in Tsukuba, Japan, and represents a major upgrade to the KEKB accelerator. It consists of two 3 km rings equipped with radiofrequency (RF) systems which accelerate electron and positron beams to 4 and 8 GeV, respectively, and make them collide in the center of the Belle~II detector. With respect to KEKB, SuperKEKB has longer dipoles and redesigned magnet lattice to squeeze the emittance, and includes additional (or modifies) RF systems for higher ($\times 2$) beam currents. SuperKEKB reduces ($\times 20$)  the beam spot size using a nano-beams scheme achieved with new superconducting final focusing  quadrupole magnets near the interaction region, and further increases the luminosity by introducing crab cavities.  
The last impart a kick to restore head-on collisions for an initial large crossing angle,  avoiding long-range beam-beam collisions. The upgraded collider  will deliver an integrated luminosity of 50~ab$^{-1}$ and reach a peak  luminosity of $8 \times 10^{35}$~cm$^{-2}$s$^{-1}$, about 50 and 40 times KEKB, respectively, meaning about $10^{10}$ $BB$ or $\tau\tau$ pairs per year. 

\subsection{The Belle II detector}
The Belle II detector consists of several subdetectors that, with respect to Belle, were upgraded or replaced for improved detection performance at higher luminosity. A larger tracker improves the impact parameter and secondary vertex resolutions, increases  the $K^0_S$ and pion efficiencies, and provides  better flavor tagging. A smaller beryllium double-wall beam pipe ($r = 1.5 \rightarrow 1.0$~cm), combined with  an innermost silicon pixel  layer much closer to the interaction region ($r = 1.4$~cm), improve the impact parameter resolution along the beam line ($\sigma_{d_z} \sim 60 \rightarrow  20~\mu $m). An upgraded time-of-propagation (TOP) counter and ring-imaging Cherenkov counters with aerogel radiator (A-RICH), together with a faster and more hermertic $K^0_L$ and $\mu$ (KLM) detector, allow a better particle identification and further enhance the flavor tagging and background rejection. Faster and more reliable trigger and data acquisition (DAQ) systems and algorithms permit the operation of Belle II at a much higher event rate.

\subsection{Tracking system}
The tracking system is formed by the vertex detector (VXD) and the central drift chamber (CDC). The VXD provides precise measurements of the impact parameter of tracks and secondary vertex of  short-lived particles, while the CDC complements  the VXD for finer track reconstruction and measurement of long-lived particles. The tracker is designed to cope with the particle multiplicity emerging from SuperKEKB collisions (consisting in average of 11 charged tracks, 5 neutral pions and 1 neutral kaon) and to measure the soft momentum of tracks.

The VXD comprises two layers of DEPFET-based pixel detectors (PXD) and four layers of double sided silicon strip detectors (SVD).
The energy loss of tracks traversing the PXD and SVD is very low; in terms of radiation lengths, the  material budget  is about 0.2\% and 0.8\% $X_0$, respectively. With about 8 million pixels, the PXD has excellent spacial granularity ($\sigma < 15 ~\mu$m); also, the excellent timing  ($2-3$~ns) of the 187 SVD sensors provides  fast readout. The VXD production is expected to be completed in 2017 and  fully installed in spring 2018.

The CDC  is made up  of 14,336 drift cells with sense wires seated in 56 concentric layers within $r = 16.8-111.1$~cm from the interaction point, forming 9 axial/stereo alternating superlayers. Axial layers are parallel to the beam line, while stereo layers are slightly twisted to enable  3-dimensional reconstruction. The CDC was completed in 2015 and commissioning with cosmics rays continues to date. Full CDC operation is expected for the 2017 data taking period.

\subsection{Particle identification system}
Two Cherenkov counter technologies exist in Belle II to  enable particle identification (PID) in the barrel (central) and endcaps (forward) regions. The TOP counter in the barrel region consists of 16 modules located in the gap between the CDC and the calorimeter. A TOP module is composed of two quartz bars instrumented with an array of photodetectors, a mirror and an extension prism that collect the Cherenkov radiation of the charged particle at the edge of the bars.
Photodetectors are micro channel plate photomultipliers (PMT) which provide ultra-fast ($\sim 50 $~ps) and hight sensitivity photon detection. 

Both, the time of propagation of the Cherenkov light throughout the TOP bar and the detection angle (PMT channel), are needed to create the space-time image that classifies between pions and kaons. In the forward region, in contrast, PID is 
obtaind directly by the
Cherenkov rings created by particles passing through the A-RICH  radiator. The last consists of two adjacent layers of aerogel with different refractive index  to increase the photon yield (by overlapping the Cherenkov rings) detected by hybrid avalanche photo detectors (HAPD).


The TOP and A-RICH counters are already built, and will be installed and operational in Belle~II by 2017.

\subsection{Calorimeter, $K^0_L$ and muon detectors}
Already installed in Belle II, an electromagnetic calorimeter (ECL) is located inside a superconducting solenoid coil that provides a 1.5~T magnetic field. The ECL uses CsI(Tl) crystals and CsI in the barrel and endcaps regions, respectively, and is equipped with faster readout electronics with respect to Belle in order  to compensate for larger backgrounds. Outside of the solenoid coil, an iron flux-return is instrumented to detect $K^0_L$ mesons and to identify muons (KLM). The KLM consists mainly of resistive plate chambers (RPC) in the barrel region and of scintillators in the endcaps region. $K^0_L$ mesons can shower hadronically in the absorptive iron plates and be detected in the KLM active detectors.

\subsection{Trigger and data acquisition systems}
Bhabha scattering and $\gamma \gamma$ processes, as well as intrabeam (Touschek)  scattering, are expected to be major sources of background
in Belle II. To deal in general with the much higher physics rate, a two-stage trigger system based on hardware (L1) and software (HLT)  will be implemented to reduce the rate to less than 10~kHz, recording only collision events of interest. The DAQ system gathers the information from the subdetectors and produces events of about 200~kB, implying a storage rate around 1.5~GB/s. Belle II makes use of a Grid computing model based on various levels, called tiers, to distribute the raw data storage,  processing, Monte-Carlo simulation production and further reprocessing to several data centers and computer cluster sites in more than 17 countries. 

\section{Physics cases}
Belle II has a rich physics program that is well documented elsewhere~\cite{Belle2note} and is currently divided in nine topics developed by corresponding working groups: (1) leptonic and semileptonic $B$ decays; (2) radiative and electroweak penguins; (3) $\phi_{1,2}$; (4) $\phi_3$; (5) charmless hadronic $B$ decays; (6) open charm decays; (7) quarkonium and quarkonium-like states; (8) $\tau$, low multiplicity  and electroweak; and (9) NP. 
 Some analyses are presented in the following sections. No attempt is made 
to describe them in detail, but only to show the discovery potential of Belle II. 


\subsection{$B \rightarrow \ell \nu_{\ell}$}
The decay of a $B^+$ ($u\bar{b}$) meson to a lepton and neutrino takes place by the exchange of a $W^+$ boson; however, it  could also be  mediated by a charged Higgs boson inspired from two-Higgs-doublet (2HDM) or Supersymmetry models~\cite{Hou, Akeroyd}. Using inputs from other decay modes and lattice QCD, it is found that $\mathcal{B}(B^+ \rightarrow \tau^+ \nu_{\tau}) = 0.76^{+0.08}_{-0.06} \times 10^{-4}$~\cite{CKMFitter}. This prediction can be modified by a factor that depends on the ratio of the Higgs doublet vacuum expectation values, normally called $\tan \beta$. Measurements of $\mathcal{B}(B^+ \rightarrow \tau^+ \nu_{\tau})$ by BaBar and Belle are consistently higher than expected, with a current average of $0.848^{+0.036}_{-0.055}\times 10^{-4}$~\cite{CKMFitter}. The  measurement by Belle~II is expected to be  more  precise than this average and, together with the decay $B^+ \rightarrow \mu^+ \nu_{\mu}$, will impose tight constrains on charged Higgs models with large $\tan \beta$ and relatively small mass $m_{H^+}$ of the charged Higgs.   

\subsection{$B \rightarrow D^{(*)} \ell \nu_{\ell}$}
\begin{figure}[h]
\begin{center}
\includegraphics[width=20pc]{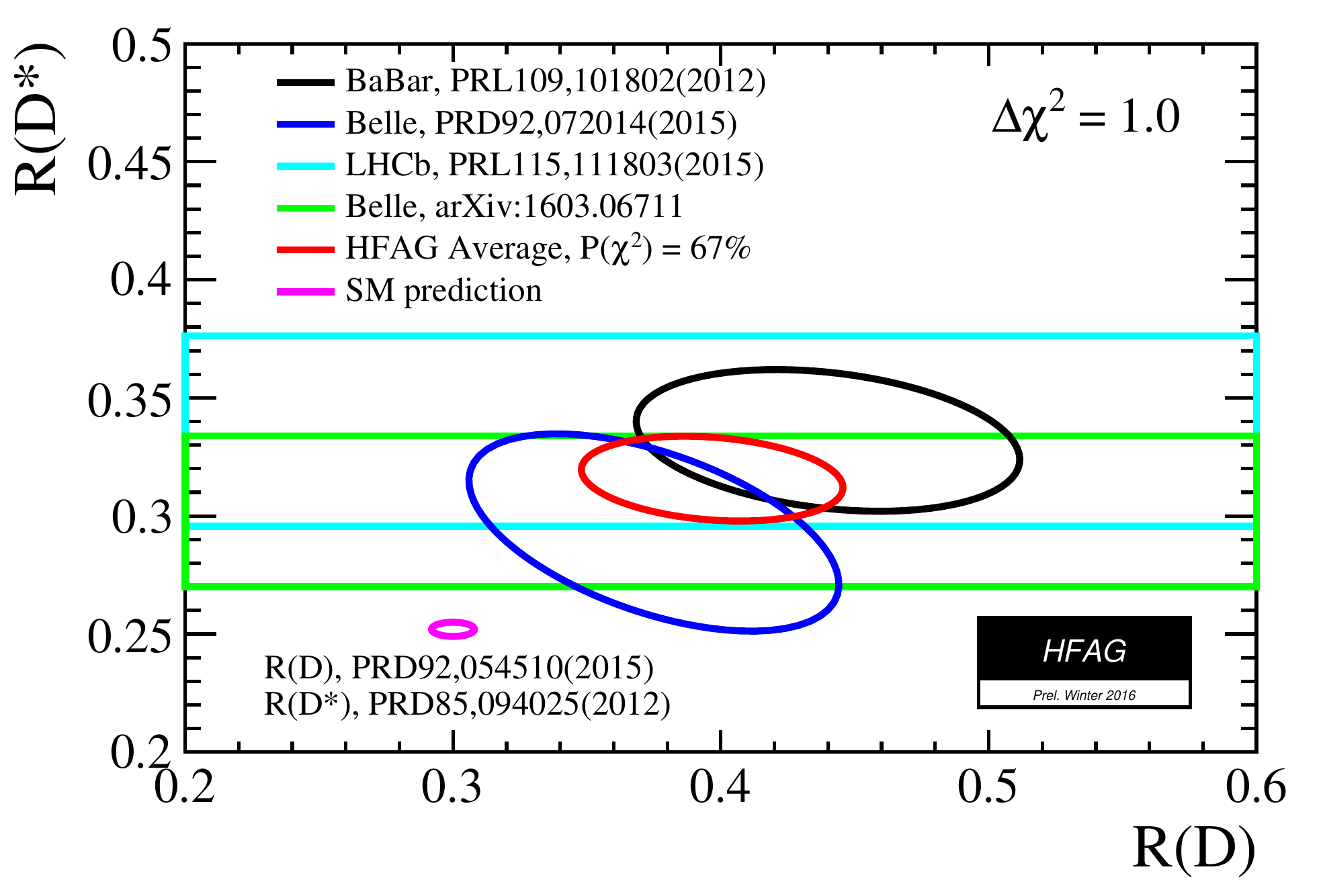}\hspace{2pc}%
\begin{minipage}[b]{14pc}\caption{\label{RRstar}SM prediction and measured values of $R(D)$ and $R(D^*)$ by BaBar, Belle and LHCb~\cite{HFAG2016RRstar}.  \vspace{0.2pc} }
\end{minipage}
\end{center}
\end{figure}

The ratios $R(D^{(*)}) = \mathcal{B}(B \rightarrow D^{(*)}\tau \nu)/\mathcal{B}(B \rightarrow D^{(*)}\ell \nu)$, where $\ell = \mu, e$, can also constrain charged Higgs models through lepton universality violation (LUV) tests. While BaBar found incompatible values of $x = \tan \beta /m_{H^+}$  for $R(D)$ and $R(D^*)$~\cite{Babar2013}, largely disfavoring the 2HDM hypothesis, Belle found agreement at $x \approx 0.5$~GeV$^{-1}$~\cite{Belle2015}. A recent measurement by LHCb~\cite{LHCb2015} is in good agreement with both BaBar and Belle results, but their combination is in disagreement  at a level $4.0\sigma$ with the SM~\cite{HFAG2016RRstar}, as shown in figure~\ref{RRstar}.  Belle II will reduce the uncertainty of these measurements by an order of magnitude, potentially resolving between the SM and 2HDM models.

\subsection{Second-class current decays}
It is worth noting that the measurement of yet undiscovered second-class currents~\cite{2ndclass} was proposed during this Workshop to improve the bounds on a charged  Higgs, certainly within reach of Belle II through $\tau^- \rightarrow \pi^- \eta^{(')} \nu_{\tau}$ decays~\cite{Roig1,Roig2}.

\subsection{Direct CPV in $B\rightarrow K\pi$ decays}
Tension in differences between CP asymmetries of $B$ decay modes, e.g. $\Delta A = A_{CP}^{B^0 \rightarrow K^+\pi^-} - A_{CP}^{B^+ \rightarrow K^+\pi^0} = -0.122 \pm 0.022$~\cite{HFAG2016ACPV}, 
could be a hint of NP  in the electroweak penguin operator, or just a result of hadronic effects. To investigate this issue, a model independent sum rule can be used to test the SM expectations~\cite{Gronau}. This requires the measurement of $A_{CP}$ in neutral final state modes (full isospin analysis), where Belle II will have  better sensitivity  than LHCb by about a factor of two.

\subsection{Unitary triangle}
Belle II will further over-constrain the unitary triangle (UT) in figure~\ref{UT} and test the CKM mechanism for CPV at the percent level, reducing the world average values of the UT angles by about a factor of four.

\subsection{Analysis of the inclusive process $b \rightarrow s \ell^+ \ell^-$}
Recent measurements by LHCb of the observables in the angular distribution of  flavor changing neutral current (FCNC) exclusive decays, $B \rightarrow X_s \ell^+ \ell^-$, present several discrepancies which can be explained by a wide range of NP scenarios~\cite{Descotes}. Either the confirmation of these anomalies or the discrimination among NP models require more precise measurements. Belle II can take advantage of its ability to reconstruct inclusive decays to perform the measurement of the inclusive branching fraction of  the process $b \rightarrow s \ell^+ \ell^-$  and the forward-backward asymmetry of the final state leptons.

\subsection{Quarkonium}
Systems composed of a quark and antiquark of the same flavor ($q\bar{q}$),  called quarkonia, have been extensively studied by Belle and many other experiments using several production mechanisms. A quarkonium is an excellent laboratory where perturbative and non-perturbative QCD models can be  tested reliably. The large data sample to be collected by Belle II represents a new opportunity to continue this research field, including  the search for and characterization of states clearly containing a $q\bar{q}$ quark pair, but with properties that deviate from theoretical expectations for quarkonium spectroscopy. Such non-conventional or ``exotic" states have been subject of extensive research in recent years (since the discovery by Belle II of the X(3872) in 2003~\cite{X3872}) and point to the existence of  tetraquarks, molecules, hybrids, etc.

\subsection{Lepton flavor violation}\label{LFV}
\begin{figure}[h]
\begin{center}
\includegraphics[width=30pc]{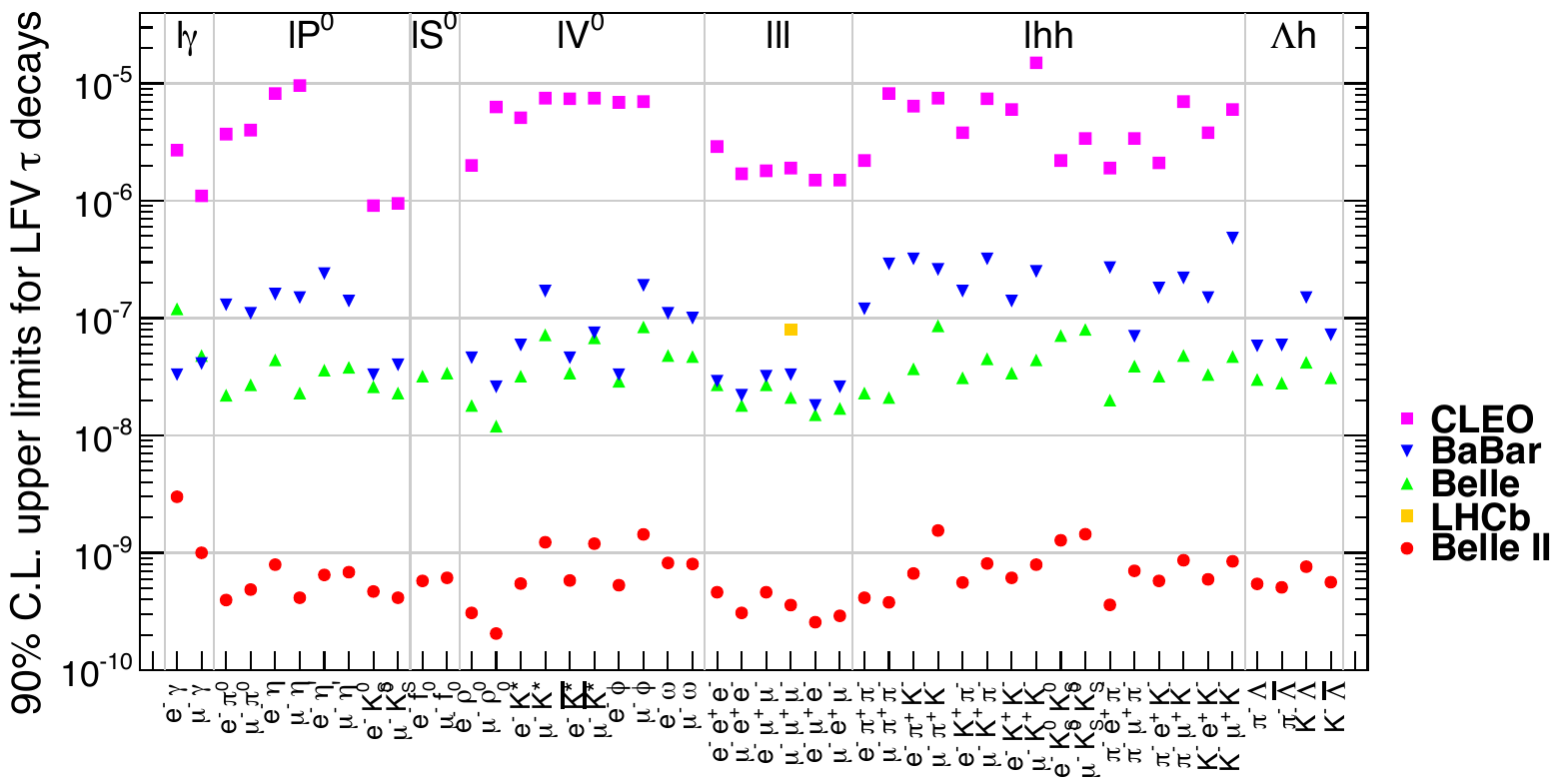}
\caption{\label{tau}Upper limits on $\tau$~LFV branching fractions~\cite{Belle2note}.
}
\end{center}
\end{figure}

Lepton flavor violating (LFV) decays
are highly suppressed in the SM and, therefore, represent a clean null test of the SM. Several NP scenarios increase the branching fraction of LFV decays to  order of magnitudes that are accesible to current and coming experiments. Belle II sensitivity for LFV decays is over 100 times greater than its predecessor  for the cleanest channels (e.g. $\tau \rightarrow 3\ell$) and about 10 times for channels where irreducible backgrounds play a major role (e.g. $\tau \rightarrow \ell \gamma$). Figure~\ref{tau} summarizes the status of searches for LFV in $\tau$ decays.

\section{The Belle II collaboration and experiment schedule}
The Belle II collaboration consist of about 650 collaborators from 100 institutions and 23 countries. A joint theory-experiment effort, called the Belle II Theory Interface Platform (B2TIP), was established to study and advance the prospects of the Belle~II physics program. 

The Belle II schedule can be divided in three phases. Beams started circulating in SuperKEKB in 2016 as part of Phase-1, with a commissioning detector, named BEAST II, to measure the beam backgrounds. During Phase-2 (2017) SuperKEKB will initiate and tune collisions, while Belle~II will operate without the VXD.  Finally, in Phase-3 (2018) the complete Belle~II detector will start taking physics data. The instantaneous luminosity is expected to evolve over time and to reach its peak value in 2022. The goal of Belle II is to collect an integrated luminosity of 50~ab$^{-1}$ by 2024.

\section{Summary}
The first generation of $B$-factories were key to establishing the CKM mechanism. In the process, several hints of NP and unexpected states were found.  In the following years, the Belle II and the  LHC experiments will complement and verify each other to find the much-expected NP. 
Belle~II is not only Belle with higher luminosity; there are also many improvements in the collision scheme, detector, trigger system, reconstruction algorithms, etc. Belle II will start operations in 2017 without the VXD, and a fully functional detector is expected in 2018 that will take data until collecting 50~ab$^{-1}$.

\ack{IHC would like to thank to Red-FAE-CONACyT and the CINVESTAV Physics Department for the funding provided to attend to the XV Mexican Workshop on Particles and Fields, and to CONACyT and SEP for the funding  provided by means of projects FOINS-296 and CB-254409.}

\section*{References}
 
 \bibliography{iopart-num}

\end{document}